\documentclass[acus]{JAC2000}

\usepackage{graphicx}

\def\degree{$^\circ$}

\begin{document}
\title{SNS FRONT END DIAGNOSTICS\thanks{Work supported by
the Director, Office of Science, Office of Basic Energy Sciences, of the U.S.
     Department of Energy under Contract No.~DE-AC03-76SF00098}}

\author{L. Doolittle, T. Goulding, D. Oshatz, A. Ratti, J. Staples,\\
E. O. Lawrence Berkeley National Laboratory, Berkeley, CA 94720, USA}

\maketitle

\begin{abstract} 

%  \footnote{stuff} OK in abstract
The Front End of the Spallation Neutron Source (SNS)
extends from the Ion Source (IS), through a 65\thinspace keV
LEBT, a 402.5\thinspace MHz RFQ, a 2.5\thinspace MeV MEBT, ending at the
entrance to the \hbox{DTL}. The diagnostics suite in this
space includes stripline beam position and phase monitors (BPM),
%
% revision from abstract-only submission: add emittance device.
%
% old:
% and toroid beam current monitors \hbox{(BCM)}.
%
% new:
toroid beam current monitors \hbox{(BCM)},
and an emittance scanner.
%
% end revision
%
Provision is included for beam profile measurement,
either gas fluorescence, laser-based
photodissociation, or a crawling wire. Mechanical and electrical design
and prototyping of BPM and BCM subsystems are
proceeding. Significant effort has been devoted
to packaging the diagnostic devices in minimal space.
Close ties are maintained to the rest of the SNS effort,
to ensure long term compatibility of interfaces
and in fact share some design work and construction.
The data acquisition, digital processing, and
control system interface needs for the BPM, BCM, and
LEBT diagnostic are similar, and we are committed to
using an architecture common with the rest of the SNS
collaboration.

\end{abstract}

\section{INTRODUCTION}

The SNS Front End consists of an H$^{-}$~Ion Source, Low Energy
Beam Transport (LEBT), a Radio Frequency Quadrupole (RFQ)
with 65\thinspace keV injection energy and 2.5\thinspace MeV output energy,
and a 3.6\thinspace m long Medium Energy Beam Transport (MEBT), that
matches and chops the 2.5\thinspace MeV H$^{-}$~beam before injection
into the remainder of the SNS linac \cite{ref-frontend}.

The extremely compact 65\thinspace keV LEBT leaves no room for conventional
diagnostics.  Only one measurement of beam properties remains,
a split-collector current measurement, that goes
under the name ``LEBT Diagnostic.''  No beam diagnostic devices
at all are included in the RFQ.

\begin{figure*}[tb]
\centering
\includegraphics*[width=170mm]{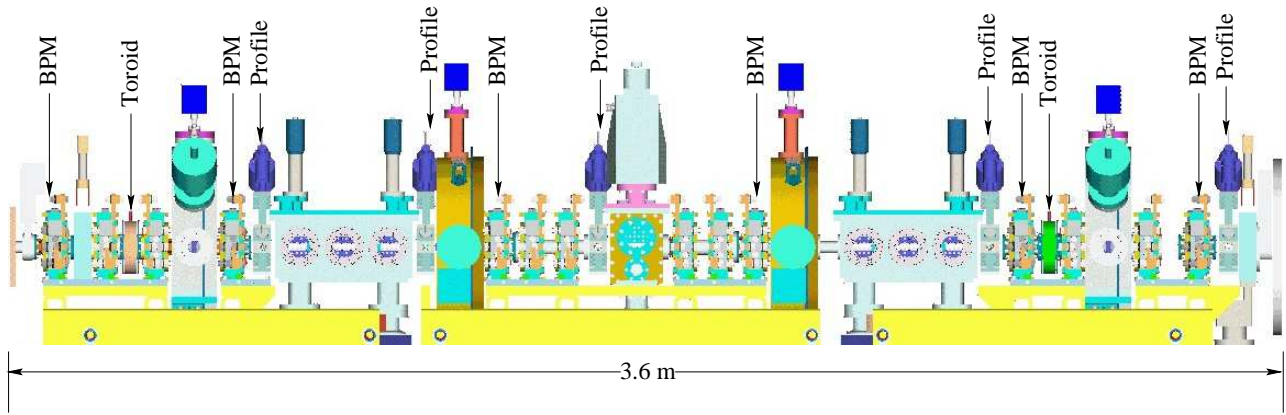}
\caption{Overview of MEBT.}
\label{fig-mebtoverview}
\end{figure*}

Table \ref{tab-summary} shows the instruments that will
be assembled on the 2.5\thinspace MeV, 3.6\thinspace m long \hbox{MEBT}.
Figure \ref{fig-mebtoverview} shows their placement along the beam line.
This paper will discuss each of these instruments in turn.
% Although all computer-generated images are shown,
% the parts are in various phases of design, construction,
% and testing, as noted individually.

% XXX can I find a way to eliminate the gap between z and extent?
\begin{table}[htb]
\begin{center}
\caption{MEBT instrumentation summary}
\begin{tabular}{|l|c|rl|l|}
\hline
\textbf{Device} & \textbf{Qty.} & \multicolumn{2}{c|}{\textbf{$z$ extent}} & \textbf{Measures}\\ \hline
LEBT      & 1 &   0&mm    & centering           \\
BPM       & 6 & 106&mm*   & position, phase     \\
BCM       & 2 &  59&mm    & current             \\
Profile   & 5 &  51&mm    & $x$ and $y$ profile \\
Emittance & 1 & 2 $\times$ 51&mm
                          & $x$-$x'$ and $y$-$y'$ \\ \hline
\end{tabular}
\break
\strut * all but 23 mm overlaps with quadrupole magnet
\label{tab-summary}
\end{center}
\end{table}

% Daryl says:
%
% Beam Box Length (emittance slit, collector, beam scraper) = 50 mm
%
% BPM Length = 106 mm (101 to 108 with bellows movement)
% Comparable beam pipe with a bellows and O-ring flanges is 83 mm long (3.25").
%
% Toroid Length = 59 mm
%
% Profile Monitor Beambox Length = 51 mm

\section{LEBT DIAGNOSTIC}

Beam current will be monitored on a four-way split electrode
(LEBT chopper target), placed at the exit of the \hbox{LEBT}.
The current balance between electrodes at different times during the chopper
cycle can be used to qualitatively determine offsets from
the RFQ axis \cite{ref-lebt-diag}.
With appropriate manipulation of the beam steering, some
information might be gained about beam size.

\begin{figure}[ht]
\centering
\includegraphics[width=45mm]{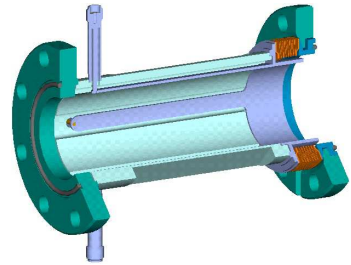}
\caption{Stripline BPM assembly}
\label{ignore1}
\end{figure}

\section{BEAM POSITION MONITORS}

BPMs will be installed in six locations in the MEBT,
spaced roughly every 90\degree\ of betatron phase advance
\cite{ref-mebt-optics}.
The BPMs will primarily be used as a secondary
standard for restoring the beam, where the primary standard is the
null point for quadrupole steering.  The BPMs also serve to measure
the trajectory of systematically deflected bunches (this pattern is
related to the betatron oscillation of particles in the bunch, but
differs due to space charge effects)
and to provide beam phase information for tuning the longitudinal
optics by way of the rebuncher cavities.
Thus, reliability, repeatability and linearity are more important than
initial zero set. 

To minimize the amount of beamline space dedicated to BPMs, the strips
are relatively narrow (22\degree) so as to fit between quadrupole pole tips.

The electrical processing will use the 805\thinspace MHz signal component,
since the fundamental 402.5\thinspace MHz signal will be contaminated by
fringe fields from nearby 402.5\thinspace MHz rebuncher cavities.

% XXX find reference and/or add a line about x^3 and x*y^2 correction
%     based on Bessel functions
Since this is a low velocity beam ($\beta=0.073$)
wire-based calibration will not give a proper calibration
curve.  A simple numerical model will convert electrical
signal strengths to linearized position.

Measurements of a prototype show the expected shorted
50\thinspace $\Omega$ stripline
behavior, with no spurious resonances below 8\thinspace GHz.
Construction of all required BPMs is nearly complete.

Electronics to measure longitudinal bunch information
now uses the signals coming from the BPM pickup,
to avoid the need for separate beamline hardware (see section 6 below).
% XXX relative -> variational ???
For relative phase measurement with a single BPM, this is
fairly easy.  For absolute measurement between pairs of BPM's,
this requires extra attention to cables and calibration.
All BPMs are installed in the same directional orientation,
so those phase signals can be compared with no additional
sensor calibration term.

\begin{figure}[ht]
\centering
\includegraphics[width=35mm]{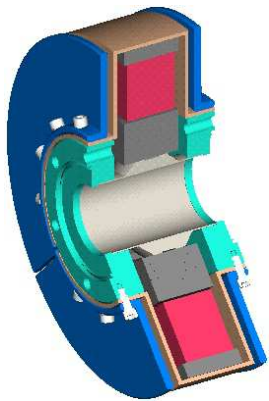}
\caption{Current transformer assembly}
\label{ignore2}
\end{figure}

\section{BEAM CURRENT MONITORS}

The MEBT beamline includes two current transformers to measure beam
current, one before and one after
the MEBT chopper target.  These will be used to measure the current
waveforms that are generated by the LEBT and MEBT chopping processes.
They also provide the first
calibrated measure of beam current and integrated beam charge.

The toroidal transformer is nearly a standard
Bergoz FCT-082-50:1 \cite{ref-bergoz},
using a high permeability core to keep droop to a minimum during
the 0.65 $\mu$s chopped beam pulse.  These transformers have a measured
droop of 0.06\thinspace \%/$\mu$s.

These devices are mounted 37\thinspace mm from the main focussing
magnet pole tips (1.16 diameter), leading to concerns that
the DC magnet fringe field would saturate portions of the
toroid core.  The result would be a increased droop rate,
and sensitivity of the measurement results on quadrupole
drive current.  Tests have shown this is indeed the case:
with the quadrupole running near its design gradient (38\thinspace T/m),
the current transformer's droop approximately doubles.
The design shown above, however, includes a 3.2\thinspace mm thick shield
made from mild steel.  With this field clamp inserted, the droop
of the transformer is not measurably affected by quadrupole operation.

The transformers have been delivered, the remaining mechanical
beamline parts have been fabricated, and assembly is underway.

\begin{figure}[ht]
\centering
\includegraphics[width=45mm]{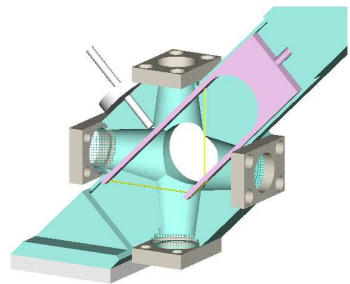}
\caption{Wire Scanner concept, with provisions for RGF or LP device.}
\label{ignore3}
\end{figure}

\section{BEAM PROFILE MONITORS}

Measurements of beam profile in the MEBT
are considered essential to check that the transverse beam optics is
behaving as intended.
In final operation of SNS, these measurements should be made
without disturbing the operation of the machine.
The two leading contenders to provide such
functionality are Residual Gas Fluorescence (RGF) \cite{ref-fluorescence}
and Laser Photodissociation (LP) \cite{ref-laserwire}.
Unfortunately, both of these techniques are considered experimental
at this time, and cannot be counted on to deliver reliable
profile data for beamline commissioning in 2002.

The current plan is to provide conventional crawling wire scanners,
with co-located optical ports, for eventual addition of an RGF or LP monitor.
The wire scanners will be used first
to commission the beamline, and then to test and commission an
optical technique.

Brookhaven Nat.\ Lab.\ will provide the flange-mounted wire scanners for
the whole SNS project,
including the \hbox{MEBT}.  That design will be customized to fit
the tight space allotment.
Unlike the final optical devices, the wire scanner
is intended to work only when the beam runs at reduced duty factor.
Rather than the full 1\thinspace ms pulse at 60\thinspace Hz, we
expect 100 $\mu$m wire to survive 100\thinspace $\mu$s
pulses at 6\thinspace Hz (1\% of the nominal 6\% duty factor).
This is adequate to commission, but not operate, the accelerator.

The beam box is designed to accept a wire scanner, plus two
pairs (in and out, $x$ and $y$) of $f$/2.8 windows on the beam,
and a gas jet that could be part of the fluorescence experiments.
Space is sufficiently tight that the beam boxes will likely be
manufactured as part of neighboring components (chopper electrodes,
chopper target, and emittance scanner).  We hope that this
optical access will be sufficient to deploy a final non-invasive
profile measurement.

\section{EMITTANCE}

The 1999 SNS Beam Instrumentation Workshop \cite{ref-1999-biw} strongly recommended that
a way be found to measure the emittance of the beam as it
leaves the MEBT on its way to the DTL.  By subsuming phase
measurement into the BPM pickup system,
a slit and multisegment collector assembly (at 51\thinspace mm each)
could be fit into the beamline.
Note that the drift space between these devices
contains one focussing quadrupole, one BPM, and one profile monitor.
The engineering design of this subsystem has started.
The slit cannot absorb the full beam power.

For each position of the movable slit, all the beam
divergence information is recorded simultaneously by
a segmented collector assembly.
Each segment has its own front-end electronics equipment,
consisting of a charge amplifier and sample-and-hold.

Table~\ref{tab-emittance} shows a plausible parameter set
for the MEBT emittance device.
With these parameters, the error in reconstructed emittance
and the error of the reconstructed Twiss beta function are
typically on the order of 2\% or less.

\begin{table}[htb]
\begin{center}
\caption{MEBT Emittance Device parameter set}
\begin{tabular}{|l l|}
\hline

Slit width                       &   0.2  mm  (7.9 mils) \\
Total slit movement range        &   5.0  mm \\
Slit positions for measurement   &   50 \\
Collector segments               &   64 \\
Collector size                   &  30   mm (square)\\
Collector center-center spacing  &  0.5  mm \\
Slit-collector spacing           &  205  mm \\
\hline
\end{tabular}
\label{tab-emittance}
\end{center}
\end{table}

\section{SIGNAL PROCESSING ELECTRONICS}

The signal processing needs for the BPM, LEBT diagnostic,
and BCM are similar, both among themselves
and with their cousins in the larger SNS project.
Collaborative and competitive development of electronics is underway
at LBNL, LANL, and BNL.

Most of the relevant information can be collected with
a moderate rate (34-68 MHz), moderate resolution (12-14 bit)
digitization of a suitably conditioned signal.  We are
investigating digitization and signal processing platforms
that can reliably and cost-effectively deal with this
volume of data, and interact with the Global Controls (EPICS).
That platform would then be used for all BPM, BCM, and LEBT
signal handling, and possibly other uses within SNS.

Each instrument has unique analog signal conditioning requirements.
For BPM processing, at least one channel of ``vector voltmeter''
is required to process beam phase information.  This is expected
to function with a mixer and direct IF sampling.
Such processing can also be used for position readout.
Log amp circuitry is also under consideration
for the actual signal strength measurements: it has a
dynamic range advantage over ordinary linear analog processing.
% XXX add more detail here maybe

The BCM signal conditioning requirements for the front end
are actually quite modest, essentially a 40\thinspace dB amplifier and filter.
This simplicity has to be balanced against compatibility with the
future BCM signal conditioning for the SNS ring, where the signal
has an additional 60 dB of dynamic range, and turn-turn differences
are important \cite{ref-bcm-electronics}.

\section{ACKNOWLEDGEMENTS}

The authors would like to thank all our collaborators at
LANL, BNL, and ORNL for their various roles in keeping
this project moving.  Contributions from
Tom Shea, John Power, Marty Kesselman, Pete Cameron, and Bob Shafer
have been particularly helpful.

\end{document}